\documentclass[12pt]{article}
\usepackage{graphicx}
\usepackage{amsmath}
\usepackage{amsfonts}
\usepackage{amssymb}

\numberwithin{equation}{section}

\begin{document}

\title{Families of conformally related asymptotically flat, static vacuum data.\\}

\author{Helmut Friedrich\\ 
Max-Planck-Institut f\"ur Gravitationsphysik\\
Am M\"uhlenberg 1\\
14476 Golm, Germany}

\maketitle

\begin{abstract}
\noindent
{\footnotesize
Extending the results of \cite{friedrich:confstatic} we give a complete description of the
asymptotically flat, conformally non-flat, static vacuum data which admit 
non-trivial, asymptotically smooth conformal mappings  onto other such data. 
These data form a $3$-parameter family which decomposes into  $1$-parameter families of data which are conformal to each other. The data and the associated static vacuum solutions are given explicitly in terms of elliptic and, in a special case, elementary functions. 
}
\end{abstract}

PACS: 04.20.Ex, 04.20.Ha, 04.20.Jb

{\footnotesize

\section{Introduction}

In a recent article  the conditions  have been analysed under which  time reflection symmetric, asymptotically flat, conformally non-flat, static vacuum data can be mapped onto other such data by conformal maps which extend smoothly to space-like infinity  (\cite{friedrich:confstatic}).  It turned out that in a certain conformal representation the Ricci tensor must be of a particular form, the data must have a non-vanishing quadrupole moment, be axi-symmetric, and admit a non-vanishing, hypersurface orthogonal,  conformal Killing near space-like infinity.  Moreover, it was shown that there exists a $3$-parameter family of data admitting conformal maps. This family exhausts the set of all such data with the possible exception of data corresponding to borderline cases which were difficult to analyse with the methods of  \cite{friedrich:confstatic}. 

To close this gap the existence problem is analysed in this article by a different method. The axial Killing field and the conformal Killing field are used to define an orthonormal frame field for the suitably rescaled metric and study the problem in the frame formalism. 
The properties of the metrics found in \cite{friedrich:confstatic} then allow  the  conformal static field equations to be reduced to three ODE's which depend on three parameters (different from the ones used earlier). The solutions to these equations can be given in terms of elliptic functions and, in a particular case, in terms of elementary functions. Their dependence on the parameters, which are different from the ones considered in \cite{friedrich:confstatic}, is such that solutions corresponding to different parameters are non-isometric. 

The $3$-parameter class of solutions is rule by curves which correspond to data that are conformal to each other. The $1$-parameter families of conformal factors relating these data to each other  are obtained by solving an ODE of the type considered before. In the new representation the borderline cases left open in  \cite{friedrich:confstatic} are easily discussed. They do not supply new solutions.
With the results of  \cite{friedrich:confstatic} it follows that the solutions are of Petrov type N on the axis and of type $I$ on an open set surrounding the axis.

Explicit expressions for some of the data and the conformal factors are given and some properties of these solutions are discussed. It turns out that the analytic extensions of data, which are related by conformal diffeomorphisms near space-like infinity, can have different global properties. The corresponding $4$-dimensional static vacuum solutions are also given explicitly.

\section{Some results on static vacuum data}

In the following we recall results obtained in \cite{friedrich:confstatic}. We refer the reader to 
\cite{friedrich:statconv} and \cite{friedrich:confstatic} for more details.
A static vacuum data set is given by  a triplet $(\tilde{S}, \tilde{h}, v)$, where $\tilde{S}$ denotes a smooth 
$3$-dimensional manifold, $\tilde{h}$ a (negative definite) metric field and $v > 0$ a function so that the 
static vacuum field
equations
\begin{equation}
\label{statEinstvac}
R_{ab}[\tilde{h}] = \frac{1}{v}\,\tilde{D}_a\,\tilde{D}_b\,v,
\quad\quad
\Delta_{\tilde{h}}\,v = 0, 
\end{equation}
hold on $\tilde{S}$ where $\tilde{D}$ denotes the covariant derivative and 
$R_{ab}[\tilde{h}]$ the Ricci tensor defined by $\tilde{h}$. We assume that these data are asymptotically flat so that there exist coordinates $\tilde{x}^a$, mapping  $\tilde{S}$ onto the complement of a closed ball in $\mathbb{R}^3$, in which 
\[
\tilde{h}_{ac} = - (1 + 2\,m\,|\tilde{x}|^{-1})\,\delta_{ac} 
+ O(|\tilde{x}|^{-(1 + \epsilon)}),
\quad 
v = 1 - m\,|\tilde{x}|^{-1} + O(|\tilde{x}|^{-(1 + \epsilon)})
\quad \mbox{as} \quad |\tilde{x}| \rightarrow \infty,
\]
with some $\epsilon > 0$. Here $|\,.\,|$ denotes the standard Euclidean norm. 
The coefficient $m$ represents the ADM mass of the data. Without loss of generality we can assume that  $m > 0$.

The set  $S = \tilde{S} \cup \{i\}$, obtained by adjoining to $\tilde{S}$ a point $i$
representing space-like infinity,
can be given a real analytic differentiable structure for which  $i$ is an inner point  to which 
the fields
\begin{equation}
\label{Omega-h-rho}
\Omega = \left(\frac{1 - v}{m}\right)^2, \quad
h = \Omega^2\,\tilde{h}, \quad
\rho = \left(\frac{1 - v}{1 + v}\right)^2,
\end{equation}
can be extended so that $\Omega \in C^2(S)$ while $h$ is a real analytic metric and $\rho$ a real analytic function on $S$ which satisfies
\begin{equation}
\label{sigmaval}
\rho = 0,\,\,\,\,\,\,D_a\rho = 0,\,\,\,\,\,\,
D_a D_b \rho = - 2\,\mu\,h_{ab}
\quad \mbox{at} \quad i, \quad
\rho > 0 \quad \mbox{on $\tilde{S}$}, \quad \quad\mu = m^2/4,
\end{equation} 
where $D$ denotes the covariant derivative defined by $h$. 
The fields $\tilde{h}$ and $v$ are obtained from $h$ and $\rho$  by 
\begin{equation}
 \label{v-hphys- from-h-rho}
v = \frac{1 - \sqrt{\rho}}{1 + \sqrt{\rho}}, \quad \quad  
\tilde{h} = \Omega^{-2}\,h \quad \mbox{with} \quad
 \Omega = \frac{\rho}{\mu\,(1 + \sqrt{\rho})^2}.
\end{equation}
The  Ricci scalar and the trace free part $s_{ab}$ of the Ricci tensor  of $h$ 
satisfy in the conformal gauge (\ref{Omega-h-rho}) 
\begin{equation}
\label{gaugechar}
R[h] = 0, \quad \quad s_{ab} = R_{ab}[h].
\end{equation}
Observing (\ref{sigmaval}), the static vacuum equations (\ref{statEinstvac}) can be written in terms of $h$ and $\rho$ in the form 
\begin{equation}
\label{n1fequ}
0 = \Sigma_{ab} \equiv D_a D_b \rho -
s\,h_{ab} + \rho\,(1 - \rho)\,s_{ab},
\end{equation}
\begin{equation}
\label{an2fequ}
2\,\rho\,s = D_a\rho\,D^a\rho
\quad \mbox{with} \quad 
s = \frac{1}{3}\,\Delta_h \rho.
\end{equation}
Equation (\ref{an2fequ}) is implied by (\ref{sigmaval})
and (\ref{n1fequ}) so that (\ref{statEinstvac}) with the asymptotic behaviour of $\tilde{h}$ and $v$
given above is in fact equivalent to (\ref{sigmaval}) and (\ref{n1fequ}).

Let $(\tilde{S}, \tilde{h}, v)$  be a static vacuum data set with ADM mass $m > 0$ and 
$(S, h, \rho)$ the corresponding conformal data.  Suppose $\gamma$, $m' $ are positive constants, 
$ \nu \equiv (m/(m'\,\gamma^2))^2$, $\mu' = m'^2/4$, and $u$ is a smooth, positive function on $S$ used to rescale $h$. 
For 
\[
\tilde{h}' =  \Omega'^{-2}\,h', \quad v' =  \frac{1 - \sqrt{\rho'}}{1 + \sqrt{\rho'}} \quad  \mbox{with} \quad  
\Omega' = \frac{\rho'}{\mu'\,(1 + \sqrt{\rho'})^2},
\] 
to define near $i$ a static vacuum data set  with ADM mass $m'$,
the  conformal rescaling of $h$  
must be complemented by a rescaling of $\rho$,
\begin{equation} \label{rescaled-fields}
h' = \left(\frac{\gamma^2\,\nu}{u}\right)^2 h, \quad \rho' = \frac{1}{u}\,\rho.
\end{equation}
The fields so obtained do define a static vacuum set if and only if the function $u$ satisfies (possibly after shrinking $S$) the conditions
\begin{equation}
\label{uati}
u(i) = \nu,
\end{equation}
\begin{equation}
\label{DDvan}
0 = \Pi_{ab} \equiv D_a D_b u - t\,h_{ab}
+ u\,(1 - u)\,s_{ab},
\end{equation}
\begin{equation}
\label{Lapvan}
0 = \Pi \equiv 2\,u\,t - D_c u D^c u 
\quad \mbox{with} \quad t = \frac{1}{3}\,\Delta_h u.
\end{equation}
If these conditions are satisfied with a function $u$ with $du \not \equiv  0$ we say that 
$(\tilde{S}, \tilde{h}, v)$  (or $(S, h, \rho)$) admits a non-trivial conformal rescaling extending smoothly to space-like infinity.

 It has also been shown  that these conditions 
can be satisfied non-trivially  if $du(i) \neq  0$ (which implies that $s_{ab}(i) \neq  0$)
and cannot be satisfied in a non-trivial way  with $du(i) = 0$ and $s_{ab}(i) = 0$
or with $du(i) = 0$ and $\nu = 1$.
Whether they can be satisfied with $du(i) = 0$, $s_{ab}(i) \neq  0$, and $\nu \neq 1$  was left open. In the following it will only be assumed that $s_{ab}(i) \neq  0$. This implies in our gauge that $h$ is conformally non-flat (\cite{friedrich:cargese}).
If $u$ satisfies (\ref{DDvan}) and (\ref{Lapvan})
a calculation shows that the general transformation law of  Ricci tensors under conformal rescalings reduces to the simple relation 
\begin{equation} 
\label{sab-transf}
s_{ab}[h'] = u\,s_{ab}[h].
\end{equation}

Suppose $(S, h, \rho)$ is conformally non-flat and admits a non-trivial conformal rescaling extending smoothly to space-like infinity. Let $U$ be an $i$-centered, convex $h$-normal neighbourhood  so that
$0 < \rho < 1$, $D_a\rho \neq 0$ on $U \setminus \{i\}$.
For the following statements to be true the set $U$ needs to (and can) be chosen small enough. 
Set 
\[
\rho_a = D_a\rho, \quad 
u_a = D_au, \quad
w = \frac{1 - u}{1 - \rho},
\quad
w_a = D_aw,
\quad 
U_* = \{w_a \neq 0\} \subset U,
\]
so that $i \in U_*$ if and only if $u_a(i) \neq 0$.
Then the following can be shown.

The set $U_*$ is dense in $U$ and there exists a  smooth function $\beta$ on
$U_*$ so that
\begin{equation}
\label{2ndbigindcond}
s_{ab}
= \beta\,(w_{a}\,w_b - \frac{1}{3}\,h_{ab}\,w_{c}\,w^c).
\end{equation}
If $V \subset  U_*$ is a connected, simply connected neighbourhood of a point
$x_* \in U_*$, there exist a constant $\beta_* \neq 0$ and a function $H =
H(w)$ defined on $V$ with $H(w(x_*)) = 0$ so that
\begin{equation}
\label{Ricexpr}
s_{ab} = \frac{\beta_*}{1 - \rho}\,e^H\,
(w_a\,w_b - \frac{1}{3}\,h_{ab}\,w_c\,w^c).
\end{equation}
$V$ can be chosen so that $\epsilon^{abc}\,u_b\,\rho_c \neq 0$ on $V$. Then there exists a
function $l = l(w) > 0$ on $V$ so that the vector field
\begin{equation}
\label{Killingfield}
X^a = l\,\epsilon^{abc}\,u_b\,\rho_c,
\end{equation}
extends to an analytic, hypersurface orthogonal Killing field $X$
on $U$ which satisfies 
\begin{equation}
\label{XrhoXuXw}
X^a \rho_a = 0, \quad X^a u_a = 0,
\end{equation}
and defines  a Killing field for $\tilde{h}$ on $U \setminus \{i\}$.
The field $X$ vanishes but $D_aX_b \neq 0$ along a certain geodesic $\gamma(\tau)$ with $\gamma(0) = i$ and  $h( \dot{\gamma},  \dot{\gamma}) = - 1$
along which $\,\dot{\gamma}^a\,D_aX_b = 0$. This geodesic defines the axis of the axi-symmetry defined by the flow of $X$.
If $du(i) \neq 0$ then $\dot{\gamma}^a(0) \sim D^au(i)$.
The integral curves of $X$ are closed near $\gamma(\tau)$. 
The field 
\begin{equation}
\label{conKillingexists}
Y^a = f\,w^a  \quad \mbox{with} \quad f = \frac{l}{l_*}\,(1 - \rho)^2,
\quad l_* = l(w(x_*)) > 0,
\end{equation}
extends to an analytic, hypersurface orthogonal conformal Killing field $Y$
satisfying
\begin{equation}
\label{XKcommute}
D_aY_b = \omega\,h_{ab} + \frac{1}{l_*}\,\epsilon_{abc}\,X^c, \quad 
h(X, Y) = 0, \quad\,\, [X, Y] = 0 \quad \mbox{on} \quad U.
\end{equation}
It is neither homothetic nor a Killing field for $\tilde{h}$.
It is tangential to the geodesic $\gamma$. If $S$ is chosen small enough, $Y \neq 0$ while
$X$ vanishes only along $\gamma$.

 It follows that $f$ is smooth and $\neq  0$ on $U_*$ while 
$|f|  \rightarrow \infty$ where $w_a = 0$. 
By (\ref{2ndbigindcond}) we can write 
\begin{equation}
\label{3rd2ndbigindcond}
s_{ab}
= \alpha\,(Y_{a}\,Y_b - \frac{1}{3}\,h_{ab}\,Y_{c}\,Y^c)
\quad \mbox{on} \quad U,
\end{equation}
where $\alpha  = \frac{3}{2}\,\frac{s_{ab}\,Y^a\,Y^b}{(Y_c\,Y^c)^2} $ is by the assumptions that $s_{ab}(i) \neq 0$ a non-vanishing smooth function.

\section{The frame field and the  coordinates}

Given a conformally non-flat data set $(S, h, \rho)$ which admits a non-trivial conformal rescaling, we use the properties discussed above to construct coordinates and an $h$-orthonormal frame field.
By the properties of $X$ and $Y$ the vector field 
\[
Z^a = \epsilon^a\,_{bc}\,X^b Y^c,
\]
is hypersurface orthogonal, vanishes on the axis, and satisfies
\begin{equation} 
\label{Zcomm's}
[X, Z] = 0, \quad \quad  [Y, Z] = \omega\,Z.
\end{equation}
A direct calculation gives on $U_* \setminus \gamma$ 
\[
<Z, d \rho>\, = \epsilon^{abc}\,\rho_a X_b Y_c
= - \frac{f}{1 - \rho}\,\epsilon^{abc}\,\rho_a X_b u_c
= - \frac{1}{l_*}(1 - \rho)\,X_c X^c  > 0.
\]
This relation extends to $U \setminus \gamma$ because both sides are analytic. It follows that 
$Z$ does not vanish on $U \setminus \gamma$ and points away from the axis.
We set
\[
p = \sqrt{- X_cX^c},
\,\,\,\quad
q = \sqrt{- Y_cY^c} > 0,\,\,\,\quad
n = \sqrt{- Z_cZ^c} = \sqrt{- X_cX^c}\,\sqrt{- Y_cY^c}
= p\,q.
\]
Because $Y$ does not vanish on $U$ the set $U$ is smoothly foliated by hypersurfaces
(the `$Y^{\perp}$-foliation')
which are orthogonal to $Y$ and thus in particular to the axis. Away from the axis these hypersurfaces are ruled by the integral curves of 
the unit vector field $\frac{1}{n}\,Z$, which approach the axis from different directions.

 If $du(i) \neq 0$  we can assume that $dw \neq 0$ on $U$  and the $Y^{\perp}$-foliation coincides with that given by the hypersurfaces $\{w = const.\}$.
 If $du(i) = 0$, $\nu \neq 1$, then $dw(i) = 0$ but we get with the expression for $D_aD_bw$ given in 
 \cite{friedrich:confstatic}
 \begin{equation}
\label{Hessw(i)}
(Y^aD_a)^2w(i) = 2\,(1 - \nu)\,\mu\,q^2 \neq 0,
\end{equation}
and for  $w_b\,w^b$ along the integral curves of $\frac{1}{n}\,Z$  the ODE   
\begin{equation}
\label{wbsqrODE}
\frac{1}{n}\,Z(w_b\,w^b) = - \frac{1}{1 - \rho}\,<\frac{1}{n}\,Z, d \rho>\,w_b\,w^b.
\end{equation}
By (\ref{Hessw(i)}) we can assume that $w_a(\gamma(\tau)) \neq 0$ for $\tau \neq 0$
so that  by (\ref{wbsqrODE}) $w_a \neq 0$ on the sets $\{w = const. \neq 1 - \nu\}$ 
which thus represent a subset of the  smooth $Y^{\perp}$-foliation. The remaining leave of this foliation, which contains $i$, then coincides with the set  $\{w = w(i) = 1 - \nu\}$. It follows from 
(\ref{wbsqrODE}) that $w_a = 0$ on this set and from (\ref{Hessw(i)})
 that the restriction of $w$ to an integral curve of $Y$ near to $i$ assumes on $\{w = w(i)\}$
 a minimum if $1 - \nu > 0$ resp. a maximum if $1 - \nu < 0$.
Thus we can assume that $U_* = U$ and $f > 0$ on $U$  if $du(i) \neq 0$ and 
$U_* = U \setminus \{w = w(i) \}$ if $du(i) = 0$.

The value of $l_*$ and thus the scaling of $X$ can and will be assumed so that there exists
on the complement of the axis a smooth function $\phi$ which coincides with a natural parameter on the integral curves of $X$ and  takes values in $[0, 2\,\pi[$, where $2\,\pi$ defines the smallest period on the integral curves (this does not fix the sense of rotation because we left the sign of $\epsilon_{abc}$ open). Then $D_aX_b = \epsilon_{abc}\,\eta^c$ with  $\eta_a \eta^a = - 1$ at $i$
and this relation is preserved along $\gamma$ because  the integrability condition for the Killing field $X$ implies
\[
D_{\dot{\gamma}}(\eta_c \eta^c) = - 1/2\,D_{\dot{\gamma}}(D_aX_b D^aX^b) =
- D^aX^b\,\dot{\gamma}^d\,X^c\,R_{cdab} = 0 \quad \mbox{along}\,\, \gamma.
\]
We define coordinates $z$ and $r$ near $i$ so that
\[
<Y, d z>\, = 1,\,\,\,
z = 0 \,\,\, \mbox{on} \,\,\, \{w = w(i)\}, \,\,\,
<Z, dr>\, = p,  \,\,\, r \rightarrow 0 \,\,\, \mbox{at the axis}.
\] 
Then
\[
q = O(1),  \,\,\, 
p = O(r), \,\,\, n = p\,q = O(r) \,\,\, \mbox{whence} \,\,
<\frac{1}{n}\,Z, dr>\, = \frac{1}{q} = O(1) 
\,\,\, \mbox{as} \,\,\, r \rightarrow 0.
\]
Since
${\cal L}_Y<X, dz>\, = \,<{\cal L}_Y\,X, dz> 
+ <X, {\cal L}_Y\,dz>\,
= \,<X, (d \circ i_Y + i_Y \circ d)\,dr>\,
= 0$ and 
${\cal L}_Y<Z, dz>\, = \,<{\cal L}_Y\,Z, dz> 
+ <Z, {\cal L}_Y\,dz>\,
= - \omega <Z, dz>$, it follows that
$<X, dz>\, = 0$ and $<Z, dz>\, = 0$ because $X$ and $Z$ are tangent to
$\{w = w(i)\}$ where $z = 0$. Thus $z = z(w)$ resp. $w = w(z)$.
Moreover, we have
$\,\,{\cal L}_Z<X, dr>\,\, = \,\,<{\cal L}_Z\,X, dr> 
+ <X, {\cal L}_Z\,dr>\,
= \,<X, d\,p>\, = 0$ 
and, by  the Killing and  the conformal Killing
equation,
\[
{\cal L}_Z<Y, dr>\, 
= - <\omega\,Z, dr> + <Y, d p>
\]
\[
= - \omega\,p + \frac{1}{p}(K^a X^b D_bX_a)
= - \omega\,p + \frac{1}{p}(- X^a X^b D_bK_a) = 0.
\] 
Because $<X, dr> \,\rightarrow 0$ and $<Y, dr> \,\rightarrow 0$ at
the axis, it follows that 
$<X, dr> \,= 0$ and $<Y, dr> \, = 0$.
Away from the axis a  smooth frame field $\{e_k\}_{k = 1, 2, 3}$ satisfying 
$h_{ij} \equiv h(e_i, e_j) = - \delta_{ij}$
 is thus given by 
 \begin{equation} 
 \label{frame-ek}
e_1 = \frac{1}{q}\,Y = \frac{1}{q}\,\partial_z,\,\,\,\quad
e_2 = \frac{1}{n}\,Z = \frac{p}{n}\,\partial_r
= \frac{1}{q}\,\partial_r,\,\,\,\quad
e_3 = \frac{1}{p}\,X = \frac{1}{p}\,\partial_\phi.
\end{equation}
In this frame the $\epsilon$-tensor
$\epsilon_{ijk}= e^a\,_i\,e^b\,_j\, e^b\,_k\,\epsilon_{abc}$ satisfies 
$\epsilon_{ijk} = \epsilon_{[ijk]}$, $\epsilon_{123} =  - 1$.
All formulas will be given from now on  in terms of the coordinates $z$, $r$, $\phi$ and the frame $e_k$ so that $i, j, k = 1, 2, 3$ denote frame indices.
Moreover, for any function $f$ (with the exception of the function $l$ for which $l_*$ has be introduced ealier) we set $f_* = f(i)$.

The unit vector field $e_1$ extends smoothly to the axis and 
coincides there with the tangent vector of  the geodesic $\gamma$ defining the axis. It holds
\[
Y^i = q\,\delta^i, \quad
- q\,\delta^1_i = 
Y_i = h(Y, e_i) = \,<f\,dw,
\frac{1}{q}\,\partial_z>\delta^1_i  = \frac{f}{q}\,w_{,z}\,\delta^1_i,
\]
which gives
\begin{equation}
\label{m-rels}
q^2 = - f\,w_{,z},
\end{equation}
whence
\begin{equation}
\label{1m-relcons}
w_{,z} < 0  \,\,\,\, \mbox{if} \,\,\,\, du(i) \neq 0, \quad
w_{,z} \rightarrow  0 \,\,\, \mbox{at} \,\,\, \{w = w(i)\} = \{z = 0\}
\,\,\,\, \mbox{if} \,\,\,\, du(i) = 0.
\end{equation}
Since, by (\ref{Hessw(i)}), 
\begin{equation}
\label{2m-relcons}
w_{,zz}(i) = 2\,(1 - \nu)\,\mu\,q^2 \neq 0 
\,\,\,\, \mbox{if} \,\,\,\, du(i) = 0,\,\,\,\, \nu \neq 1,
\end{equation}
it follows that $f$ and $w_{,z}$ must satisfy $sign(f) = - sign(w_{,z})$ and change sign on $\{z = 0\}$.

The properties of $X$, $Y$, and $Z$ imply 
\begin{equation}
\label{e2-m}
l_*\,e_2(q) = l_*\,\frac{1}{n}\,D_Zq = - p,
\end{equation}
\begin{equation} 
\label{e1-m-p}
e_1(q) = \frac{1}{q}\,D_Y q =  \omega,
\quad \quad
e_1(p) = \frac{1}{q}\,D_Y p = \frac{\omega\,p}{q},
\end{equation}
whence
\begin{equation}
\label{gammadefprop}
e_1\left(\frac{p}{q}\right) = 0
\quad \mbox{so that} \quad 
\psi \equiv \frac{p}{q} 
\quad \mbox{depends only on $r$}.
\end{equation}
With the $1$-forms
\[
\sigma^1 = q\,dz, \quad
\sigma^2 = q\,dr, \quad
\sigma^3 = p\,d\phi,
\]
dual to the frame fields $e_k$ the metric can thus be written
\begin{equation}
\label{h-in-Weyl-form}
h = - \delta_{ik}\,\sigma^i\,\sigma^k 
= - q^2\,\{dz^2 + dr^2 + \psi^2\,d\phi^2\}.
\end{equation}
Using the invariance of  the static field equations under rescalings of the metric with constant conformal factors  we assume in the following, without loosing generality, that 
\[
q_* = 1,
\]
and  thus also that $\gamma = 1$ in (\ref{uati}), (\ref{rescaled-fields}). 

Since $q$ is independent of $\phi$, $X = \partial_{\phi}$ is clearly a Killing field.
The form of the metric  implies that $Y = \partial_z$ is a conformal Killing field for $h$ because it is a Killing field for the metric in curly brackets conformal to $h$.
The function $\psi$ behaves  at the axis as follows.
Since  $X = 0$ there, the fields $\rho_k$, $u_k$, $w_k$ and $\eta_k$ must all be tangent to the axis and thus $\eta^k = \epsilon\,\delta^k\,_1$ with $|\epsilon| = 1$. This implies that
$e_i\,(p) = - \frac{1}{p}\,X^k\,D_iX_k \rightarrow  - \epsilon\,\epsilon_{i31} = \epsilon\,\delta^2_i$
as $r \rightarrow 0$ along an integral curve of $e_2$.
Because $e_2$ is pointing away from the axis and $p = 0$ on while $p > 0$ away from the axis 
we must have $\epsilon = 1$. It follows that
\begin{equation}
\label{psi-as-r-0}
\psi = r + O(r^2) \quad \mbox{near the axis}.
\end{equation}
The Ricci tensor assumes by (\ref{3rd2ndbigindcond}) the form
\begin{equation}
\label{sij}
s_{ij} = \zeta\,p_{ij}
\quad \mbox{with} \quad \zeta_* \neq 0, \quad
p_{ij} = \delta^1\,_i\,\delta^1\,_j + \frac{1}{3}\,h_{ij},
\end{equation}
where $\zeta$ denotes a smooth function. 
The dualized Cotton tensor
$B_{ij} =  \frac{1}{2}B_{ikl}\,\epsilon_j\,^{kl} =
\frac{1}{2}\,D_{k}(R_{li} - \frac{1}{4}R\,h_{li})\,\epsilon_j\,^{kl}$
is given by (cf.  \cite{friedrich:confstatic})
\begin{equation}
\label{Cotton}
B_{ij} = \frac{1}{l_*}\,\zeta\,\psi\,\,\delta^1\,_{(i}\,\delta^3\,_{j)}.
\end{equation}

The connection coefficients in the frame (\ref{frame-ek}), defined by $D_{e_k}e_j = \Gamma_k\,^l\,_j\,e_l$ and  satisfying $\Gamma_{ijk} = - \Gamma_{ikj}$
with $\Gamma_{ijk} = \Gamma_i\,^l\,_k\,h_{lj}$, are given by
\[
\Gamma_l\,^i\,_j = 0 \quad \mbox{if} \quad l \neq i \neq j \neq i,
\quad \quad
\Gamma_1\,^1\,_3 = 0, \quad
\Gamma_2\,^2\,_3 = 0, \quad
\]
\begin{equation}
\label{KZXGammas}
\Gamma_1\,^1\,_2 = A, \quad
\Gamma_2\,^2\,_1 = B, \quad
\Gamma_3\,^3\,_1 = C, \quad 
\Gamma_3\,^3\,_2 = D,
\end{equation}
with
\[
A = \frac{e_2(q)}{q}, \quad
B = \frac{e_1(q)}{q}, \quad
C = \frac{e_1(p)}{p}, \quad 
D = \frac{e_2(p)}{p}. 
\]
The Bianchi identity for (\ref{sij}) thus reads in our frame
\[
0 = D^k\,s_{kj}  = - e_1(\zeta)\,\delta^1\,_j + \frac{1}{3}\,e_j(\zeta)
+ \zeta\,(\Gamma_1\,^1\,_j -  h^{ik}\,\Gamma_i\,^1\,_k\,\delta^1\,_j)
\]
\[
= \frac{\zeta}{3}\left\{\delta^1\,_j\,e_1(\log(\zeta^2\,p^3\,q^3)) 
+ \delta^2\,_j\,e_1(\log(|\zeta|\,q^3)) \right\}.
\]
 In view of (\ref{gammadefprop}) this is equivalent to
\begin{equation}
\label{sij-Bianchi-id}
\zeta\,q^3 = \zeta_* \quad \mbox{with} \quad 
 \zeta_* \neq 0.
 \end{equation}
 The Ricci tensor is given by
\[
R_{jl} =
\left(- e_1(B) - e_2(A) - e_1(C)  - A^2  - B^2  - C^2  
- A\,D \right)\,\delta^1_j\,\delta^1_l
\]
\[
+ \left(- e_1(B) - e_2(A) - e_2(D) - A^2  - B^2  - D^2
 - B\,C\right)\,\delta^2_j\,\delta^2_l
\]
\[
+ \left(- e_2(D) - e_1(C)  - C^2  - D^2 - A\,D
- B\,C\right) \delta^3_j\,\delta^3_l
+ \,U\,(\delta^2_j\,\delta^1_l  + \delta^1_j\,\delta^2_l),
\]
\[
\mbox{with} \quad U \equiv - e_1(D) + A\,C - C\,D
= - e_2(C) + B\,D - C\,D.
\]

\section{The static field equations}

With (\ref{gammadefprop}) and the form of the Ricci tensor above, we find 
(\ref{sij}) to be  equivalent to
\begin{equation}
\label{ric11}
- \frac{2}{3}\,\zeta = \frac{1}{2}(- s_{11} + s_{22} +
s_{33}) = - e_2(D) - D^2 - B\,C =
- \frac{e_2(e_2(p))}{p} - \frac{e_1(q)\,e_1(q)}{q^2}, 
\end{equation}
\begin{equation}
\label{ric22}
\frac{1}{3}\,\zeta = \frac{1}{2}(s_{11} - s_{22} + s_{33})
= - e_1(C) - C^2 - A\,D
= - \frac{e_1(e_1(q))}{q}
- \frac{e_2(q)\,e_2(p)}{p\,q},
\end{equation}
\begin{equation}
\label{ric33}
\frac{1}{3}\,\zeta = \frac{1}{2}(s_{11} + s_{22} - s_{33})
= - e_1(B) - e_2(A) - A^2 - B^2 =
- \frac{e_1(e_1(q))}{q}
- \frac{e_2(e_2(q))}{q},
\end{equation}
\begin{equation}
\label{s12vanishing}
0 = U = - e_2(C) + B\,D - C\,D
= - e_2\left(\frac{e_1(q)}{q}\right) = \frac{1}{q}
\left(\frac{1}{q}\right)_{,zr}.
\end{equation}
This hyperbolic equation is somewhat surprising in the present context. It does not fix the solution but implies an important structural property, namely
\begin{equation}
\label{m-is-lambda+xi}
\frac{1}{q} = \xi(z) + \lambda(r), \quad \quad 
\xi_* + \lambda_* = 1,
\end{equation}
with certain functions  $\xi(z)$ and $\lambda(r)$. 
Subtracting (\ref{ric33}) from (\ref{ric22}) gives with 
(\ref{gammadefprop}) 
\[
0 = e_2(e_2(q))
- e_2(q)\left(\frac{e_2(\psi)}{\psi} 
+ \frac{e_2(q)}{q}\right)
= - \left(\frac{1}{q}\right)_{,rr} 
+ \left(\frac{1}{q}\right)_{,r}\,\frac{\psi_{,r}}{\psi},
\]
and thus with  (\ref{m-is-lambda+xi})
\[
\left(\frac{1}{\psi}\,\left(\frac{1}{m}\right)_{,r}\right)_{,r} 
=
\left(\frac{1}{\psi}\,\,\lambda_{,r}\right)_{,r} = 0.
\]
This is essentially (\ref{e2-m}), which allows us with (\ref{gammadefprop}), 
(\ref{m-is-lambda+xi}) to write
\begin{equation} 
\label{gamma-m-lambda}
\psi = - l_*\,\frac{e_2(q)}{q}  = l_*\,\lambda_{,r} .
\end{equation}
It follows that 
\begin{equation}
\label{e2e2m}
p\,e_2(e_2(q)) =
e_2(q)\,e_2(p),
\end{equation}
and equations (\ref{ric22}) and (\ref{ric33}) are seen to be  identical. We will ensure later that 
(\ref{ric11}) -  (\ref{ric33}) will be satisfied.

The relations $<X, dw>\, = 0$, $<Z, dw>\, = 0$  imply that $w = w(z)$ and thus 
\begin{equation}
\label{u-form}
u = 1 - w(z)\,(1 - \rho).
\end{equation}
With (\ref{conKillingexists}) and  the last of equations (\ref{m-rels}) we get 
$q = (1 - \rho)\,\sqrt{- \frac{l}{l_*\,w_{,z}}}$. Because $l = l(w)$ this can be written
with some functions $Q = Q(z)$ and $\tau = \tau(z)$ so that $Q = \xi - \tau$  in the form  
\begin{equation}
\label{rho-form}
\rho = 1 - Q\,q
= 1 - (\xi - \tau)\,q = \frac{\tau + \lambda}{\xi + \lambda}.
\end{equation}

We study now (\ref{n1fequ}).  The relations $m_{,\phi} = 0$ and (\ref{s12vanishing}) imply  that $\Sigma_{ij} = 0$ for $i \neq j$.  The equations which remain to be considered read
\[
0 = 3\,\Sigma_{jj} = 3\,e_j(e_j(\rho))  - e_1(e_1(\rho))  - e_2(e_2(\rho)) 
+ 3\,\rho\,(1 - \rho)\,s_{jj}
\]
\[
-  (3\,\Gamma_j\,^1\,_j - \Gamma_2\,^1\,_2 - \Gamma_3\,^1\,_3)\,e_1(\rho)
+ (\Gamma_1\,^2\,_1 + \Gamma_3\,^2\,_3 - 3\,\Gamma_j\,^2\,_j)\,e_2(\rho). 
\]
Observing here the connection coefficients given above, (\ref{e2e2m}), and the first of expressions  (\ref{rho-form}), which implies
\[
e_1(\rho) = - Q_{,z} - Q\,e_1(q), \quad e_1(e_1(\rho))= - e_1(Q_{,z}) - e_1(Q)\,e_1(q)
 - Q\,e_1(e_1(q)),
\]
\[
e_2(\rho) = - Q\,e_2(q), \quad  e_2(e_2(\rho)) = - Q\,e_2(e_2(q)),
\]
whence  by (\ref{e2e2m}) also
\[
\frac{e_2(p)}{p} \,e_2(\rho) = e_2(e_2(\rho)),
\]
one finds that  $\Sigma_{11} = - 2\,\Sigma_{22} = - 2\,\Sigma_{33}$ with
\[
3\,\Sigma_{33} = - e_1(e_1(\rho))  + e_2(e_2(\rho)) 
+ \frac{e_1(q)}{q}\,e_1(\rho)
-  \frac{e_2(q)}{q} \,e_2(\rho)
- \rho\,(1 - \rho)\,\zeta
\]
\[
= \frac{1}{q}\,Q_{,zz}+ Q\left(e_1(e_1(q)) 
-  \frac{e_1(q)}{q}\,e_1(q)
- e_2(e_2(q)) +  \frac{e_2(q)}{q} \,e_2(q)\right)
\]
\[
- (1 - Q\,q)\,Q\,q\,\frac{\zeta_*}{q^3}
= 
- \xi\,\tau_{,zz}  
+ \tau\,\xi_{,zz}
- \zeta_*\,\xi^2\,\tau
+  \zeta_*\,\xi\,\tau^2
\]
\[
+ \lambda\,(
\xi_{,zz} - \tau_{,zz}
-  \zeta_*\,\xi^2
+ \zeta_*\,\tau^2)
+ (\xi - \tau)\,(\lambda_{,rr}
-  \zeta_*\,\lambda^2),
\]
where  the second of the representations (\ref{rho-form}) has been used in the last step.
Since $z$ and $r$ are independent variables, the equation $\Sigma_{33} = 0$  can  hold with non-constant functions $\xi$, $\tau$, $Q$, $\lambda$ 
 if and only if there exist constants $\kappa$, $\kappa_1$ so that
\begin{equation}
\label{xi-equ}
\xi_{,zz}  -  \zeta_*\,\xi^2 - \kappa\,\xi  =  \kappa_1, 
\end{equation}
\begin{equation}
\label{tau-equ}
\tau_{,zz}   -  \zeta_*\,\tau^2    - \kappa\,\tau =  \kappa_1, 
\end{equation}
\begin{equation}
\label{lambda-equ}
\lambda_{,rr} -  \zeta_*\,\lambda^2 + \kappa\,\lambda = \kappa_1.
\end{equation}

The initial data for the functions $\xi$, $\tau$, and $\lambda$ and the values of the constants $\kappa$, $\kappa_1$, $l_*$ are determined as follows. The functions $q = \frac{1}{\xi + \lambda}$ and  
$\rho = 1 - \frac{\xi - \tau}{\xi + \lambda}$ are  not affected by transitions $\xi \rightarrow \xi - a$, $\tau \rightarrow \tau - a$,  $\lambda \rightarrow \lambda + a$ with $a \in \mathbb{R}$.
The new fields will again satisfy the equations above (and below) if  the  
constants $\kappa$, $\kappa_1$ (and the constants of integrations entering the first integrals below) are 
transformed  appropriately. This leaves the freedom to specify $\xi_*$. 
Since nothing is gained by keeping this freedom while the following choice renders the equations in a concise form, we set 
\begin{equation}
\label{xi*=1}
\quad  \xi_* = 1  \quad \mbox{and thus } \quad \lambda_* = 0.
\end{equation}
By (\ref{psi-as-r-0}), (\ref{gamma-m-lambda}) we must assume
\begin{equation}
\label{lambda-ders}
(\lambda_{,r})_* = 0, \quad (\lambda_{,rr})_* = 1/l_*. 
\end{equation}
Conditions  (\ref{sigmaval}) are then satisfied iff
\begin{equation} 
\label{tauzz-lambdarr}
\tau_* = 0, \quad
(\tau_{,z})_* = 0, \quad 
(\tau_{,zz})_* = (\lambda_{rr})_* = 2\,\mu > 0.
\end{equation}
These relations imply with (\ref{tau-equ}) resp. (\ref{lambda-equ}) 
\begin{equation}
\label{l*-kappa1}
l_* = 1/(2\,\mu), \quad
\kappa_1 = 2\,\mu. 
\end{equation}
Since  (\ref{tauzz-lambdarr}) imply that $\tau_{,z} \not\equiv 0$, $\lambda_{,r} \not\equiv 0$ we get the first integrals
\begin{equation}
\label{C-lambda-equ}
\lambda^2_{,r} 
- \frac{2}{3}\,\zeta_*\,\lambda^3 + \kappa\,\lambda^2 - 4\,\mu\,\lambda = 0,
\end{equation}
\begin{equation}
\label{C-tau-equ}
\tau_{,z}^2 - \frac{2}{3}\,\zeta_*\,\tau^3 - \kappa\,\tau^2 
- 4\,\mu\,\tau = 0.
\end{equation}
If  $\xi_{,z} \not \equiv 0$, we get the further first integral
\begin{equation}
\label{C-xi-equ}
\xi_{,z}^2 - \frac{2}{3}\,\zeta_*\,\xi^3 - \kappa\,\xi^2
- 4\,\mu\,\xi = \kappa_2.
\end{equation}
The value of $\kappa_2$ is determined  as follows. With (\ref{gamma-m-lambda}) 
and the equations above a direct calculation gives
\[
\frac{e_2(e_2(p))}{p} + \frac{e_1(q)\,e_1(q)}{q^2} =  \frac{2}{3}\,\zeta
+ (\epsilon_{\xi} - 1)\left(\frac{2}{3}\,\zeta_*\,\xi^3 + \kappa\,\xi^2 + 4\,\mu\,\xi \right)
+ \epsilon_{\xi}\,\kappa_2,
\]
\[
\frac{e_1(e_1(q))}{q} + \frac{e_2(p)\,e_2(q)}{p\,q} = - \frac{1}{3}\,\zeta
+ (\epsilon_{\xi} - 1)\left(\frac{2}{3}\,\zeta_*\,\xi^3 + \kappa\,\xi^2 + 4\,\mu\,\xi \right)
+ \epsilon_{\xi}\,\kappa_2,
\]
where $\epsilon_{\xi} = 0\,$ if $\,\xi = const. = 1\,$ and $\epsilon_{\xi} = 1\,$ otherwise.
To satisfy equations (\ref{ric11}), (\ref{ric22}), whence also (\ref{ric33}), and to 
be able to solve (\ref{xi-equ})  and (\ref{C-xi-equ}), we thus assume 
\begin{equation}
\label{Ahatkappa3-kappa3}
 \quad \kappa_2 = 0, \quad \quad  
 \frac{2}{3}\,\zeta_* + \kappa\ + 4\,\mu \ge 0 \quad
 \mbox{if} \quad \xi_{,z} \not\equiv 0, \quad \quad 
\end{equation}
\begin{equation}
\label{Bhatkappa3-kappa3}
 \zeta_* + \kappa  + 2\,\mu = 0, \quad 
 \quad \frac{2}{3}\,\zeta_* + \kappa\ + 4\,\mu = 0
\quad  \mbox{if} \quad \xi \equiv 1.\quad \quad \quad \quad \quad \,\,\,
\end{equation}

The solutions we are seeking are uniquely characterized by the data and the equations above.
As seen in the following, equations (\ref{C-lambda-equ}) to (\ref{C-xi-equ}) can be explicitly integrated  in terms of elliptic and, in a limiting case, of elementary functions. 

\subsection{The $1$-parameter classes of conformal data.}

We assume that a solution to the ODE problems above is given
and study whether equations (\ref{uati}), (\ref{DDvan}), (\ref{Lapvan})
can be solved  on this background  in a non-trivial way. With (\ref{u-form}) and (\ref{rho-form}) we can write 
\begin{equation}
\label{u-expression}
u = 1 + \chi\,q \quad \mbox{with} \quad \chi(z) = w(z)\,(\tau(z) - \xi(z)).
\end{equation}
A direct calculation involving (\ref{m-is-lambda+xi}) and equations
(\ref{xi-equ}), (\ref{tau-equ}), (\ref{lambda-equ}),  (\ref{C-lambda-equ}),  (\ref{C-tau-equ}), (\ref{C-xi-equ})
gives, whether $\xi_{,z}$ vanishes identically or not,
\begin{equation}
\label{Pi-jk-expl}
\Pi_{jk} = \frac{1}{q}
\left\{ (\chi + \xi)_{,zz} - \zeta_*\,(\chi + \xi)^2 - \kappa\,(\chi + \xi)
- 2\,\mu  \right\} p_{jk},
\end{equation}
\begin{equation}
\label{Pi-expl}
\Pi =
- \frac{2}{3}(\chi + \xi + \lambda)
\left\{ (\chi + \xi)_{,zz} - \zeta_*\,(\chi + \xi)^2 - \kappa\,(\chi + \xi)
- 2\,\mu  \right\}
\end{equation}
\[
+ (\chi + \xi)_{,z}^2 - \frac{2}{3}\,\zeta_*\,(\chi + \xi)^3
- \kappa\, (\chi + \xi)^2 - 4\,\mu\, (\chi + \xi).
\]
The non-trivial solutions to (\ref{uati}), (\ref{DDvan}), (\ref{Lapvan})  are thus given 
by (\ref{u-expression}) 
where $\chi = \bar{\chi} -  \xi$, $\xi$ given by the background, and $\bar{\chi}$
solves
\begin{equation}
\label{2nd-order-chi-bar-equ}
\bar{\chi}_{,zz} - \zeta_*\,\bar{\chi}^2 - \kappa\,\bar{\chi}
= 2\,\mu, \quad \quad
\bar{\chi}_*  = \nu, 
\end{equation}
\begin{equation}
\label{1st-order-chi-bar-equ}
\bar{\chi}_{,z}^2 - \frac{2}{3}\,\zeta_*\,\bar{\chi}^3
- \kappa \bar{\chi}^2 - 4\,\mu\, \bar{\chi} = 0,
\end{equation}
with a given constant  $\nu > 0$. 

The set of $\nu > 0$ satisfying
$\frac{2}{3}\,\zeta_*\,\nu^3 + \kappa \nu^2 + 4\,\mu \nu \ge 0$ contains besides 
$\nu = 1$  certainly all sufficiently small values $\nu > 0$ because $\mu > 0$.
For $\nu$ in this set we use (\ref{1st-order-chi-bar-equ})
 to determine $(\bar{\chi}_{,z})_*$ (with some choice of the sign).
This determines a unique solution  $\bar{\chi}$ to 
(\ref{2nd-order-chi-bar-equ}).
 If  $\bar{\chi}_{,z} \not \equiv 0$, the third equation represents  a first integral. If 
$\bar{\chi} = const.$, we must have had  $(\bar{\chi}_{,z})_* = 0$ and 
the third equation will be satisfied because it holds  at $z = 0$.

It can be seen now that the basic properties of the metric coefficients  and of the function $\rho$ are preserved under the rescalings with the conformal factors $u = u_{\nu}$. Denoting the solution $\bar{\chi}$ satisfying $\bar{\chi}_* = \nu$ by 
$\bar{\chi}_{\nu}$, the transformed fields can be written
\[
h_{\nu}  = (\frac{\nu}{u_{\nu}})^2\,h = - q_{\nu}^2(dz^2 + dr^2 + \psi_{\nu}^2\,d\phi^2), \quad \quad 
\rho_{\nu} = \frac{1}{u_{\nu}}\,\rho = \frac{\tau_{\nu}+ \lambda_{\nu}}{\xi_{\nu} + \lambda_{\nu}},
\]
with
\[
q_{\nu} = \frac{\nu}{u_{\nu}}\,q = \frac{1}{\xi_{\nu} + \lambda_{\nu}}, \quad \psi_{\nu} = \psi,
\]
where the functions
\[ 
\xi_{\nu} =  \frac{1}{\nu}\,\bar{\chi}, \quad
\tau_{\nu} = \frac{1}{\nu}\,\tau, \quad 
\lambda_{\nu} = \frac{1}{\nu}\,\lambda,
\]
satisfy the initial conditions and the equations above  with the constants
\begin{equation}
\label{const-nu-transf}
\zeta_{* \nu} = \zeta_*\,\nu, \quad \kappa_{\nu} = \kappa,  \quad 
\mu_{\nu} = \mu/\nu, \quad l_{*\nu} = 1/(2\,\mu_{\nu}) = \nu\,l_*.
\end{equation}
The  first of these relations reflects (\ref{sab-transf}), the third one
has been discussed in \cite{friedrich:confstatic}, and the last one justifies
with (\ref{gamma-m-lambda}) that we set $\psi_{\nu} = \psi$.
The corresponding $4$-dimensional static vacuum solutions are given by  
\[
\tilde{g}_{\nu} = \left(\frac{1 - \sqrt{\rho_{\nu}}}{1 + \sqrt{\rho_{\nu}}}\right)^2 dt^2 
- \frac{\mu^2(1 + \sqrt{\rho_{\nu}})^4}{\rho_{\nu}^2}\,q_{\nu}^2\,(dz^2 + dr^2 + \psi^2\,d\phi^2),\quad \quad
v_{\nu} = \frac{1 - \sqrt{\rho_{\nu}}}{1 + \sqrt{\rho_{\nu}}}.
\]

To understand the effect of choosing the sign of $(\bar{\chi}_{,z})_*$ in the discussion above, we note that the function $\tau$ is an even functions of $z$, because the ODE of second order and the initial data for $\tau$ are invariant under the coordinate reflection $z \rightarrow -z$. The transformed fields obtained for different signs of $(\bar{\chi}_{,z})_*$ are thus isometric because they are related by this reflection. It follows that up to this reflection  the rescaling reproduces the original metric if $\nu = 1$.

The borderline cases in which the quadrupole moment $\frac{m}{2}\,s_{jk}(i)$ does not vanish but the differential of the conformal factor $u$ vanishes at $i$ represented singular cases 
and remained untouched in \cite{friedrich:confstatic}.
In the present setting they are easy to discuss and it turns out that different situations can occur.
By  (\ref{u-expression}) the condition $D_ku(i) = 0$ is equivalent to $(\bar{\chi}_{,z})_* =  \nu\,(\xi_{,z})_*$.  If $\xi \equiv 1$, this requirement  implies that $(\bar{\chi}_{,z})_* = 0$ and (\ref{Bhatkappa3-kappa3}) implies $\zeta_*  = 6\,\mu$ and $\kappa = - 8\,\mu$. Equation  (\ref{1st-order-chi-bar-equ}) can then only be satisfied at $z = 0$ in the trivial case $\nu = 1$. If $\xi_{,z} \not \equiv 0$ the requirement $(\bar{\chi}_{,z})_* =  \nu\,(\xi_{,z})_*$  is seen with equations (\ref{C-xi-equ}) and (\ref{1st-order-chi-bar-equ}) to be equivalent to $\zeta_*\,\nu = 6\,\mu$. 
This excludes the case where $\zeta_* < 0$.
Given $\zeta_*$, $\kappa$, $\mu$ satisfying the inequality in (\ref{Ahatkappa3-kappa3}) and  $6\,\mu \neq \zeta_* > 0$, the condition will be met non-trivially with $\nu = 6\,\mu/\zeta_*$ and the appropriate choice of sign in solving (\ref{1st-order-chi-bar-equ}). Because the condition will not be met with the `inappropriate' sign it follows that up to isometries the solutions found in this article agree with the ones discussed in 
\cite{friedrich:confstatic}. 

\subsubsection{ The independence and interpretation of the parameters.}

Any static initial data set is characterized near space-like infinity uniquely (up to rotations of the frame at $i$) by its mass and  its null data
 (\cite{friedrich:statconv}), which are given in the present gauge by the trace free symmetric parts of the covariant derivatives of $s_{kl}$ at $i$. 
While the latter provide in general 
$2\,p + 5$ independent coefficients at order $p$,  in the case of axi-symmetry there is only one  coefficient  free at each order.
In the present case we have the mass $m$ and the first two null data are given by
$s_{kl}(i) = \zeta_*p_{kl}$ (apart from a factor the quadrupole moment, the dipole moment vanishes in the given conformal gauge)  and 
\[
(D_{(i}s_{jk)})(i) = 
\zeta_*\,
\sqrt{\frac{2}{3}\,\zeta_* + \kappa + 4\,\mu}
\,\,\{
2\,\delta^1\,_i\,\delta^1\,_j\,\delta^1\,_k - 3\,\delta^1\,_{(i}\,\delta^2\,_j\,\delta^2\,_{k)}
- 3\,\delta^1\,_{(i}\,\delta^3\,_j\,\delta^3\,_{k)}\},
\]
(essentially the octopole moment).
While $\mu > 0$, $\zeta_* \neq 0$, and $\kappa$ are restricted by inequalities,  the solutions are genuinely dependent on these parameters. 
This $3$-parameter set decomposes into $1$-parameter classes of data 
which are conformally related to each other by the conformal factors $u_{\nu}$.
The relations (\ref{const-nu-transf}) show that the rescalings considered above yield in general metrics which are not isometric to the original ones.
The examples discussed below show that the rescaled and unrescaled spaces can differ substantially.

\section{Some explicit solutions.}

In the following we integrate some of the solutions and discuss some of their properties. 

\vspace{.2cm}

\noindent
{\it The case $\zeta_* > 0$, $9\,\kappa^2 = 96\,\mu\,\zeta_*$}. The solutions are obtained in terms of elementary functions.
Denoting derivatives by a dot, equations
(\ref{C-lambda-equ}), (\ref{C-tau-equ}), (\ref{C-xi-equ}), (\ref{1st-order-chi-bar-equ})
read
\begin{equation}
\label{universal-x-equ}
\dot{x}^2 = \frac{2}{3}\,\zeta_*\,x\, \left(x + \delta\,A\,\right)^2 ,
\quad \quad A = 3\,|\kappa|/(4\,\zeta_*) = \sqrt{6\,\mu/\zeta_*},
\quad \quad \delta = \epsilon\,sign(\kappa),
\end{equation}
where
\begin{equation}
\label{epsilon-kappa}
\epsilon = 1 \quad \mbox{if} \quad 
x =\xi, \, \tau \,\,\mbox{or} \,\,\bar{\chi}
, \quad \quad  
\epsilon = -1 \quad \mbox{if} \quad 
x = \lambda.
\end{equation}
With $y = \sqrt{x/A}$ one obtains (with a choice of sign) the easily integrated equation 
\[
\dot{y} = c\,(y^2 + \delta) \quad \mbox{with} \quad c = \sqrt{\zeta_*\,A/6} = \sqrt{|\kappa|/8}.
\]
We only consider the case $\kappa < 0$. Then
\[
\xi = A\,
\left(\frac{\sqrt{A}\,\tanh(c\,z) - 1}{\tanh(c\,z) - \sqrt{A}}\right)^2, \quad \quad
\tau = A\,\tanh^2(c\,z), \quad \quad   \lambda = A\,\tan^2(c\,r),
\]
\[
\bar{\chi}_{\nu}  = A\,\left(\frac{\sqrt{A}\,\tanh(c\,z) - \sqrt{\nu}}{\sqrt{\nu}\,\tanh(c\,z) - \sqrt{A}}\right)^2.
\]
There is no restriction on $\nu$.
The  metrics $h_{\nu} = - q_{\nu}^2(dz^2 + dr^2 + \psi^2\,d\phi^2)$  in the conformal class are given by
\[
q_{\nu} = 
\frac{\nu}{A\,\left(\frac{\sqrt{A}\,\tanh(c\,z) - \sqrt{\nu}}{\sqrt{\nu}\,\tanh(c\,z) - \sqrt{A}}\right)^2
+ A\,\tan^2(c\,r)}, 
\quad \quad
\psi = \frac{1}{c}\,\frac{\sin (c\,r)}{\cos^3 (c\,r)},
\]
while the metrics $\tilde{h}_{\nu} = \Omega_{\nu}^{-2}\,h_{\nu}$ and the potentials $v_{\nu}$ can be calculated with
\[
\rho_{\nu} = \frac{\tanh^2(c\,z) + \tan^2(c\,r)}
{\left(\frac{\sqrt{A}\,\tanh(c\,z) - \sqrt{\nu}}{\sqrt{\nu}\,\tanh(c\,z) - \sqrt{A}}\right)^2 +  \tan^2(c\,r)}.
\]
With the particular choice $\nu = A$  the functions $q_{\nu}$ and $\rho_{\nu}$ reduce  to 
\[
q = q_A =  \cos^2(c\,r), \quad \quad
\rho = \rho_A = 1  - \frac{\cos^{2}(c\,r)}{\cosh^{2}  (c\,z)}.
\]
This is the one and only case in which $\xi_{,z} \equiv 0$. It follows that $Y$ is a Killing field for $h \equiv h_A$ (but not for $\tilde{h} \equiv \tilde{h}_A$ or for $h_{\nu}$, $\nu \neq A$.).

The metric $h$ is smooth and non-degenerate  on 
$M = \{ z \in \mathbb{R},\,\,0 \le c\,r < \pi/2\}$,
the function $\rho$ is smooth on $M$, vanishes only at the point $i$ given by $z = 0$, $r = 0$, and $\rho \rightarrow 1$ as  $c\,r \rightarrow  \pi/2$.
The fields $v$, $\Omega$ are smooth and positive and  the `physical' 3-metric 
$\tilde{h} = \Omega^{-2}\,h$ is smooth and non-degenerate on $M \setminus \{i\}$.
But
\[
q \rightarrow 0, \quad \psi \rightarrow \infty, \quad
v|_{\{z = 0\}} \rightarrow 0 \quad  \mbox{as} \quad 
c\,r \rightarrow  \pi/2,
\]
and the invariant $|B[h]|_h^2 \equiv B_{ij}[h]\,B^{ij}[h] =  \zeta^2\,\psi^2/(2\,l_*^2)$, obtained from (\ref{Cotton}),
is strongly divergent as $c\,r \rightarrow  \pi/2$. Since $\Omega$ assumes a
 finite positive limit as $c\,r \rightarrow  \pi/2$ and  $|B[h]|_h$ is a conformal density, it follows 
 that $|B[\tilde{h}]|_{\tilde{h}}$ also diverges as $c\,r \rightarrow  \pi/2$. 
 
 The behaviour of  $h_{\nu}$, $\nu \neq A$, is not much different as $c\,r \rightarrow  \pi/2$.
The regions where $z \rightarrow \pm \infty$ do not define asymptotically flat ends for any $\nu > 0$ because outside the axis $\{r = 0\}$, along which $B_{ij}[\tilde{h}_{\nu}] = 0$, the invariant 
$|B[\tilde{h}_{\nu}]|_{\tilde{h}_{\nu}}$ approaches positive values as $z \rightarrow \pm \infty$.

If $\nu > A$, consider the hypersurface  $H = \{z = \hat{z}\}$ with $\tanh (c\,\hat{z}) = \sqrt{A/\nu}$. The fields $q_{\nu}$  and $\rho_{\nu}$ vanish on $H$, the function $v_{\nu}$ assumes the value $1$ 
and tensor $s_{ab}[h_{\nu}]$ and $B_{ij}[h_{\nu}]$ diverge there. The fields $\tilde{h}_{\nu}$ and 
$v_{\nu}$ extend, however,  analytically across $H$ and $v_{\nu}$ grows unboundedly as 
$z \rightarrow \infty$. 

The case $\nu < A$ is more interesting.
The point $i_{\nu}$ with coordinates $r = 0$ and  $z = z_{\nu}$ with
$\tanh (c\,z_{\nu}) = \sqrt{\nu/A}$ is of particular interest because $\bar{\chi}(i_{\nu}) = 0$. It follows   
\[
q_{\nu} \rightarrow \infty, \quad \rho_{\nu}  \rightarrow \infty, \quad v_{\nu} \rightarrow - 1,
\quad \Omega_{\nu} \rightarrow 1/\mu_{\nu} \quad \mbox{as} \quad 
(z, r) \rightarrow (z_{\nu}, 0),
\]
so that neither $h_{\nu}$ nor $\tilde{h}_{\nu}$ extends smoothly to $i_{\nu}$. Rescaling 
$\tilde{h}_{\nu}$ with  the function 
\[
\bar{\Omega}_{\nu} = ((1 + v_{\nu})/m_{\nu})^2 = (\sqrt{\mu_{\nu}}(1 + \sqrt{\rho_{\nu}}))^{- 2},
\]
we get the metric  
\[
\bar{\Omega}_{\nu}^2 \,\tilde{h}_{\nu} =
- \left(\frac{\nu}{\tanh^2(c\,z) + \tan^2(c\,r)}\right)^2
\left\{dz^2 + dr^2 + \psi^2\,d\phi^2\right\},
\]
which extends smoothly to $i_{\nu}$. The function $\bar{\Omega}_{\nu}$ vanishes quadratically at $i_{\nu}$ but its Hessian with respect to the metric in curly brackets does not vanish and it  is in fact proportional to that metric. It follows that $i_{\nu}$ represents a further space-like 
infinity for the metric $\tilde{h}_{\nu}$. 
On the hypersurfaces $H_{\pm} = \{z = z_{\pm}\}$, 
$\tanh(c\,z_{\pm}) = \sqrt{A/\nu} \pm \sqrt{A/\nu - 1}$, holds 
$v = 0$ and $dv \neq 0$. Since then $Hess_{h_{\nu}}v = 0$ on $H_{\pm}$  by the static field equations, it follows   that these hypersurfaces are totally geodesic. Since $0 < z_- < z_{\nu} < z_+$, they separate  the infinities $i$ and $i_{\nu}$.
We note that $i_{\nu}$ and $H_{\pm}$ are shifted to $`z = \infty'$ as 
$\nu \rightarrow A$ and do not exist for $\tilde{h}_A$ while they are shifted to $i$ 
and $\{z = 0\}$ respectively as $\nu \rightarrow 0$.

It follows in particular that the manifolds underlying the analytic extensions of the rescaled and the original solutions need not be diffeomorphic.

\vspace{.2cm}

\noindent
{\it The case  $\zeta_* > 0$, $\,\,9\,\kappa^2 > 96\,\mu\,\zeta_*$.}
 In the following we shall need results on Jacobi's elliptic function
$sn\,(z, k)$, $cn\,(z, k)$, $dn\,(z, k)$
with (fixed) `modulus' $k$ and  `complementary
modulus' $k' $, satisfying  $k^2 + k'^{2} = 1$ and $0 \le k < 1$, $0 < k' \le 1$. 
Considered as functions on the real line the functions above have periods
$4\,K$, $4\,K$, and $2\,K$ respectively, where $K = K(k) > 0$ is given by
the `complete elliptic integral of the first kind'.
The reader if referred to \cite{lawden} for the properties of elliptic functions used in the following.

We choose now $a = R + i\,I \in \mathbb{C}$ with real numbers $R = R_{\epsilon}$ and $I = I_{\epsilon}$ such that 
$a^2 = - (3\,\epsilon\,\kappa - i\,\sqrt{96\,\mu\,\zeta_* - 9\,\kappa^2})/(4\,\zeta_*)$.
Denoting derivatives by a dot, the independent variable by $s$,  and assuming again (\ref{epsilon-kappa}), equations
(\ref{C-lambda-equ}), (\ref{C-tau-equ}), (\ref{C-xi-equ}), (\ref{1st-order-chi-bar-equ})
then read
\[
\dot{x}^2 = \frac{2}{3}\zeta_*\,x\,(x - a^2)\,(x - \bar{a}^2).
\]
It follows that  we must have $x \ge 0$. With $y = \sqrt{x}$ and a choice of sign  the equations read
\[
\dot{y} = \sqrt{\zeta_*/6}\,\sqrt{S_+\,S_-} \,\,\, \mbox{with} \,\,\,
 S_{\pm} = \frac{M \pm R}{2\,M}\,(y + M)^2 + \frac{M \mp R}{2\,M}\,(y - M)^2, 
\,\,\, M = \sqrt{R^2 + I^2}.
\]
Setting $\kappa = (\sqrt{\zeta_*/6}(M + R))^{-1}$, we find that the function 
\[
f(s) = \sqrt{\frac{M + R}{M - R}}\,\,\frac{y(\kappa\,s) - M}{y(\kappa\,s) + M},
\]
satisfies 
\[
\dot{f} = \sqrt{(1 + k'^2\,f^2)(1 + q^2)} \quad \mbox{with} \quad
k' = k'_{\epsilon} = \frac{M - R}{M + R}.
\]
This is the equation satisfied by Jacobi's elliptic function
\[
sc (u, k) = \frac{sn(u, k)}{cn(u, k)} \quad \mbox{with modulus} \quad
k = k_{\epsilon} = \frac{2\,\sqrt{M\,R}}{M + R},
\]
so that $f(s) = sc(\pm s + s_0, k)$ with some constant $s_0$. From this one gets 
\[
x(s) = M^2 \left(
\frac{1 + \sqrt{\frac{M - R}{M + R}}\,sc\left(\sqrt{\zeta_*/6}\,(M + R)\,(\pm s + s_0), k\right) }
{1 - \sqrt{\frac{M - R}{M + R}}\,sc\left(\sqrt{\zeta_*/6}\,(M + R)\,(\pm s + s_0), k\right) }
\right)^2.
\]
Adjusting the constant $s_0$ so as to satisfy the respective initial conditions, the functions $\xi$, $\tau$, $\lambda$, and $\bar{\chi}_{\nu}$ (with no restriction on $\nu$) and thus the functions $\psi$, $q_{\nu}$, $\rho_{\nu}$ and the field $h_{\nu}$, $\tilde{h}_{\nu}$, $v_{\nu}$ can be determined. It turns out that these solutions also have curvature singularities.
The properties of these solutions will not be analyzed any further here.
The case  $\zeta_* > 0$, $\,\,9\,\kappa^2 < 96\,\mu\,\zeta_*$ can be discussed similarly. 

\vspace{.2cm}

\noindent
{\it The case $\zeta_* < 0$.} This case is somewhat more interesting because the curvature of the solutions remains bounded.
We write equations
(\ref{C-lambda-equ}), (\ref{C-tau-equ}), (\ref{C-xi-equ}), (\ref{1st-order-chi-bar-equ})
in the form
\begin{equation}
\label{universal-x-equ}
\dot{x}^2 = -\frac{2}{3}\,\zeta_*\,x\, (x + a^2 )\,(b^2 - x ),
\end{equation}
with real constants  $a, \,b > 0$ satisfying
\[
a^2 = \frac{- 3\,\epsilon\,\kappa + \sqrt{9\,\kappa^2 - 96\,\mu\,\zeta_*}}{- 4\,\zeta_*},
\quad \quad 
b^2 = \frac{3\,\epsilon\,\kappa + \sqrt{9\,\kappa^2 - 96\,\mu\,\zeta_*}}{- 4\,\zeta_*},
\]
where again (\ref{epsilon-kappa}) is assumed.
The right hand side of (\ref{universal-x-equ}) is non-negative and consistent with 
the initial conditions  only if $b \ge 1$, which is equivalent to the inequality 
required in (\ref{Ahatkappa3-kappa3}),
and  if $x \ge 0$. Assuming this, we set
\[
c = \sqrt{-\zeta_*/6}, \quad
e =  \frac{\sqrt{a^2 + b^2}}{a\,b} = \left(\frac{3\,\kappa^2 - 32\,\mu\,\zeta_*}{48\,\mu^2}\right)^{1/4},
\quad
\mbox{so that} \quad  a\,b\,c = \sqrt{\mu},
\]
\[
k_{\epsilon} = \frac{b}{\sqrt{a^2 + b^2}}, 
\quad k'_{\epsilon}  = \frac{a}{\sqrt{a^2 + b^2}},  \quad \quad
k \equiv k_+ = k'_-, \quad \quad k' \equiv k_- = k'_+.
\]

Consider the cases $x =\xi, \, \tau \,\,\mbox{or} \,\,\bar{\chi}$.
It the function $f$ satisfies $f(\sqrt{\mu}\,e\,(z + z_0)) = e\,\sqrt{x(z)}$ with a number $z_0$ to be determined later,  equation (\ref{universal-x-equ}) transforms into
\[
\dot{f} = \sqrt{(1 + k^2\,f^2)(1 - k'^2\,f^2)}.
\]
This is the differential equation satisfied by Jacobi's elliptic function 
\[
sd\,(z, k) = \frac{sn\,(z, k)}{dn\,(z, k)}.
\]
It is analytic on the real line, has period $4\,K$, and satisfies $sd(z + 2\,K(k), k) =  - sd(z, k)$. 
Observing the initial conditions, it follows that 
\[
\tau = \frac{1}{e^2}\,sd^2( \sqrt{\mu}\,e\,z, k),  \quad \quad
\bar{\chi}_{\nu} =  \frac{1}{e^2}\,sd^2( \sqrt{\mu}\,e\,(z + z_{\nu}), k).
\]
The numbers $z_{\nu}$ are chosen so that 
$sd( \sqrt{\mu}\,e\,z_{\nu}, k) = \sqrt{\nu}\,\,e$, where it is assumed that   $1, \nu \in ]0,  b^2]$, $z_{\nu}$ is a continuous function of $\nu$, and, for definiteness, that 
$0 < z_{\nu} \le K/( \sqrt{\mu}\,e)$. The numbers are then determined uniquely because
(\ref{Ahatkappa3-kappa3}) ensures that $b \ge 1$ and
$sd$ is strictly increasing on the interval $[0, K(k)]$ with minimum $sd(0, k) = 0$ and maximum 
$sd(K(k), k) = 1/k' = b\,e$.
We note that $z_{\nu}$ is a strictly increasing function of $\nu$ with $z_{\nu} \rightarrow 0$ as $\nu \rightarrow 0$ and $z_{b^2} = K/( \sqrt{\mu}\,e)$. The function $\xi$ is given by 
$\bar{\chi}_1$. In a similar way one gets
\[
\lambda = \frac{1}{e^2}\,sd^2( \sqrt{\mu}\,e\,r, k').
\]
The metrics $h_{\nu}$, $\tilde{h}_{\nu}$ and the function $v_{\nu}$  are obtained from 
\[
q_{\nu} = \frac{\nu\,e^2}{sd^2( \sqrt{\mu}\,e\,(z + z_{\nu}), k) + sd^2( \sqrt{\mu}\,e\,r, k')}, 
\]
\[
\psi = l_*\,\lambda_{,r} = 
 \frac{1}{ \sqrt{\mu}\,e}\,\frac{sn( \sqrt{\mu}\,e\,r, k')\,\,cn( \sqrt{\mu}\,e\,r, k')}{dn^3( \sqrt{\mu}\,e\,r, k')},
\]
\[
\rho_{\nu} = \frac{sd^2( \sqrt{\mu}\,e\,z, k) + sd^2( \sqrt{\mu}\,e\,r, k')}
{sd^2( \sqrt{\mu}\,e\,(z + z_{\nu}), k) + sd^2( \sqrt{\mu}\,e\,r, k')},
\]
which reduce for $\nu = 1$ to the functions $q$, $\psi$, $\rho$ defining $h$, $\tilde{h}$ and $v$.

\vspace{.2cm}

\noindent
{\it Some properties of these solutions}

The functions $\xi$, $\tau$, $\bar{\chi}_{\nu}$, $\lambda$, $q_{\nu}$, $\psi $, and $\rho_{\nu}$ are periodic in $z$ and $r$ with periods $2\,K(k)/(\sqrt{\mu}\,e)$ and $2\,K(k')/(\sqrt{\mu}\,e)$ respectively.
Since $\psi = 0$ at $r = 2\,K(k')/(\sqrt{\mu}\,e)$,  the coordinate circle 
$r = 2\,K(k')/(\sqrt{\mu}\,e)$ for given $z$ must thus be identified to a point and $r$ be restricted to
the range $0 \le r \le 2\,K(k')/(\sqrt{\mu}\,e)$. The resulting manifold is diffeomeorphic to
$\mathbb{R} \times S^2$.

On the curve $r = 2\,K(k')/(\sqrt{\mu}\,e)$ so obtained 
the vector field $X$ vanishes and the curve represents a second axis for the flow of $X$ with neighbourhoods isometric to suitable neighbourhoods of the axis through $i$. 
The periodicity in $r$ implies that $\rho_{\nu} = 0$ at the point $i_{\star}$ on this axis at which
$z = 0$. It  thus represents another space-like infinity. 
The periodicity in $z$ implies that $\rho_{\nu} = 0$ at the points with
$z = 2\,K(k)\,j/(\sqrt{\mu}\,e)$, $j \in \mathbb{Z}$, and $r = 0$
or $r = 2\,K(k')/(\sqrt{\mu}\,e)$. They represent space-like infinities whose coordinate location is independent of $\nu$.

At the points with $z =  2\,K(k)\,j/(\sqrt{\mu}\,e) \, - \,z_{\nu}$,  $j \in \mathbb{Z}$ ,
and $r = 0$ or $r = 2\,K(k')/(\sqrt{\mu}\,e)$ the functions $q_{\nu}$ and  $\rho_{\nu}$ have poles. The conformal factor $\Omega_{\nu}$ has positive limits there while $v_{\nu} \rightarrow - 1$.
If  the conformal  factor $\bar{\Omega}_{\nu} 
= ((1 + v_{\nu})/m_{\nu})^2 = \mu_{\nu}^{-1}\,(1 + \sqrt{\rho_{\nu}})^{-2}$,  which vanishes quadratically at these points,   is used  to  rescale $\tilde{h}_{\nu}$, we get the metric $\bar{\Omega}_{\nu}^2\,\tilde{h}_{\nu} = \rho_{\nu}^{-2}\,h_{\nu}$, which extends smoothly to these points.  These points thus represent further spatial infinities, whose location depends on $\nu$.

The sets $H_j = \{z = K\,j/(\sqrt{\mu}\,e)\, - \,z_{\nu}/2\}$, defined by the equations 
$sd( \sqrt{\mu}\,e\,(z + z_{\nu}), k) = \pm sd( \sqrt{\mu}\,e\,z, k)$, seperate the domains
of positive and negative $v_{\nu}$. Since $v_{\nu} = 0$, $dv_{\nu} \neq 0$ on the hypersurfaces $H_j$, they are totally geodesic. 
The hypersurfaces closest to $i$, $i_{\star}$  are $H_0$ and $H_1$. 
No infinities except $i$ and $i_{\star}$ are lying in between $H_0$ and $H_1$. The  poles of $q_{\nu}$ and $\rho_{\nu}$ closest to $i$, $i_{\star}$ respectively  
have coordinates $z_- = - z_{\nu}$ and $z_+ = 2\,K(k)/(\sqrt{\mu}\,e) - z_{\nu}$. 

As $v \rightarrow b^2$ the sets $H_0$, $H_1$  approach the sets
$\{z = \mp K(k)/(2\,\sqrt{\mu}\,e\,)\}$ respectively.  In the limit they are located symmetrically  with respect to $i$, $i_{\star}$ and so are the poles at $z = z_{\mp}$.
If $\nu \rightarrow 0$, the set $H_1$ approaches 
$\{z = K(k)/(\sqrt{\mu}\,e)\}$ and $z_+ \rightarrow 2\,K(k)/(\sqrt{\mu}\,e)$. 
The set $H_0$ and the poles at $z = z_-$ approach, however, the infinities $i$ and  $i_{\star}$. It is clear that there does not exist a regular limit. 

Unfortunately, the apparent regularity of the solution is spoiled by a further identification which needs to be made. The function $\psi$ also vanishes at
$r = K(k')/(\sqrt{\mu}\,e)$. For given value of $z$  the coordinate circle $r = K(k')/(\sqrt{\mu}\,e)$  thus represents  metrically one point. If the corresponding identification
resulted in a smooth Riemannian metric  on
$M = \{0 \le r \le K(k')/(\sqrt{\mu}\,e)\}$, the set $\{r = K(k')/(\sqrt{\mu}\,e)\}$ would represent another axis for the flow of $X$. The geodesics starting at a point $x_{\circ}$ of this axis orthogonally to it in the direction of decreasing $r$ would then generate 
a set  $\Sigma_{\circ}$ which represented a smooth $2$-surface
ruled by the circles tangent to $X$. Let $x_{\bullet}$ be a point on one of these geodesics and denote by  $s_{\bullet}$ the length of the geodesic arc connecting $x_{\circ}$ with $x_{\bullet}$. The length of the circle through $x_{\bullet}$ generated by $X = \partial_{\phi}$ is then given by $L_{\bullet} = \int \sqrt{h(X, X)}\,d\phi = 2\,\pi\,p(x_{\bullet})$ and thus
\[
\lim _{x_{\bullet} \rightarrow x_{\circ}} L_{\bullet}/s_{\bullet} = - 2\,\pi\,e_2(p)|_{x_{\circ}} =
- 2\,\pi\,\partial_r\,\psi(x_{\circ}) = 2\,\pi\,sd^2(K(k'), k') = 2\,\pi/k^2 > 2\,\pi.
\]
It follows that $\Sigma_{\circ}$ has a conical singularity at $x_{\circ}$ so that  the space 
$(h, M)$ violates the requirement of elementary flatness along the line $\{r = K(k')/(\sqrt{\mu}\,e)\}$.

}

\end{document}